# Scaling laws and simulation results for the self–organized critical forest–fire model

S. Clar, B. Drossel, and F. Schwabl
*Institut für Theoretische Physik,*
*Physik–Department der Technischen Universität München,*
*James–Franck–Str., D–85747 Garching, Germany*

(April 25, 1994)

We discuss the properties of a self–organized critical forest–fire model which has been introduced recently. We derive scaling laws and define critical exponents. The values of these critical exponents are determined by computer simulations in 1 to 8 dimensions. The simulations suggest a critical dimension $d_c = 6$ above which the critical exponents assume their mean–field values. Changing the lattice symmetry and allowing trees to be immune against fire, we show that the critical exponents are universal.

PACS numbers: 05.40.+j, 05.70.Jk, 05.70.Ln

## I. INTRODUCTION

Some years ago, Bak, Tang, and Wiesenfeld introduced the *sandpile model* which evolves into a critical state irrespective of initial conditions and without fine tuning of parameters [1]. Such systems are called *self–organized critical* (SOC) and exhibit power–law correlations in space and time. The concept of SOC has attracted much interest since it might explain the origin of *fractal structures* and $1/f$*–noise*. Models for earthquakes [2,3], the evolution of populations [4,5], the formation of clouds [6] and river networks [7], and many more have been introduced and investigated by computer simulations, but in contrast to the sandpile model, most of these SOC models are barely understood. For most of them exists no proof that they really are critical, and little analytic treatment has been presented so far. It even has been conjectured that systems without conservation laws cannot be SOC.

This conjecture has been refuted when a forest–fire model without conservation laws was shown to be critical under the condition that time scales are separated [8]. In one dimension, where the model is still nontrivial, the exact values of the critical exponents have been calculated [9], thus proving the criticality of the model. A scaling theory presented in [8] suggested classical values for the critical exponents in two and more dimensions but has been shown to be too simple [10–12]. Scaling laws for the static properties of the model have been given in [12]. The values of several critical exponents for the two–dimensional model have been determined by computer simulations [10,12]. These simulations differ in the values of two of the critical exponents. Simulations in up to 6 dimensions [11] suggest that the critical behavior of the forest–fire model becomes percolation–like above 6 dimensions.

In this paper, we present a scaling theory for the statics and dynamics of the system and present simulation results in 1 to 8 dimensions. Additional critical exponents and scaling relations are introduced, and the values of the critical exponents are determined by computer simulations, which are performed closer to the critical point than the earlier simulations. The universality of the values of the critical exponents is investigated by changing the lattice symmetry and by considering the case of nonvanishing immunity.

The outline of the paper is as follows: In Sec. II, the rules of the model are introduced and the origin of the SOC behavior is explained. In Sec. III, the scaling theory of the model is presented. In Sec. IV, simulation results in 1 to 8 dimensions are given, and the critical behavior in high dimensions is discussed. Sec. V investigates the universality of the critical exponents. In the final section, the results are summarized and discussed.

## II. THE MODEL

The forest–fire model is a stochastic cellular automaton which is defined on a $d$–dimensional hypercubic lattice with $L^d$ sites. Each site is occupied by a tree, a burning tree, or it is empty. During one time step, the system is parallely updated according to the following rules

- burning tree $\longrightarrow$ empty site
- tree $\longrightarrow$ burning tree, if at least one nearest neighbor is burning
- tree $\longrightarrow$ burning tree with probability $f$, if no neighbor is burning
- empty site $\longrightarrow$ tree with probability $p$.

An even more general forest–fire model also contains an immunity $g$ [13,14]. In its original version, introduced by P. Bak, K. Chen, and C. Tang, the forest–fire model contained only the tree growth parameter $p$ [15]. This version of the model shows regular spiral–shaped fire fronts in the limit of slow tree growth [16,17]. Throughout this paper, we will assume that the system size $L$ is large enough such that no finite–size effects occur. In the simulations, we have always chosen periodic boundary conditions.

Starting with arbitrary initial conditions, the system approaches after a transition period a steady state the properties of which depend only on the parameter values. Let $\rho_e$, $\rho_t$, and $\rho_f$ be the mean density of empty sites, of trees, and of burning trees in the steady state. These densities are related by the equations

$$\rho_e + \rho_t + \rho_f = 1 \tag{2.1}$$

and

$$\rho_f = p\rho_e . \tag{2.2}$$

The second relation says that the mean number of growing trees equals the mean number of burning trees in the steady state. Large–scale structures and therefore criticality can only occur when the fire density $\rho_f$ is small. When the fire density is large, trees cannot live long enough to become part of large forest clusters. Eq. (2.2) shows that the fire density becomes small when the tree growth rate $p$ approaches zero. In addition, the lightning probability $f$ must satisfy

$$f \ll p\,. \tag{2.3}$$

Otherwise a tree is destroyed by lightning before its neighbors grow, and no large–scale structures can be formed.

These conditions are not yet sufficient to bring about critical behavior in the forest–fire model. When lightning strikes a small forest cluster, it burns down very fast, before any tree can grow at its edge. But when lightning strikes a large forest cluster, it needs some time to burn down, and new trees might grow at the edge of this cluster while it is still burning so that the fire is never extinguished. In order to observe critical, i.e. self–similar behavior, small and large forest clusters must burn down in the same way. We therefore choose the tree growth rate $p$ so small, that even the largest forest cluster burns down rapidly, before new trees grow at its edge. In this case, the dynamics of the system depend only on the ratio $f/p$, but not on $f$ and $p$ separately. When $f$ and $p$ are both decreased by the same factor, the overall time scale of the system is also changed by this factor, but not the number of trees that grow between two lightnings and therefore not the size distribution of forest clusters and of fires. The condition that forest clusters burn down rapidly can be written in the form

$$p \ll T^{-1}(s_{\max})\,, \tag{2.4}$$

where $T(s_{\max})$ is the time the fire needs to burn down a large forest cluster and will be determined below (see Eq. (3.20)). In the simulations, this condition is most easily realized by assuming that forest clusters burn down instantaneously, i.e. during one time step. This simplified version of the SOC forest–fire model [8] has been mentioned first in [18].

The inequalities (2.3) and (2.4) represent a *double separation of time scales*

$$T(s_{\max}) \ll p^{-1} \ll f^{-1}, \tag{2.5}$$

which is the condition for SOC behavior in the forest–fire model. The time in which a forest cluster burns down is much shorter than the time in which a tree grows, which again is much shorter than the time between two lightning occurrences. Separation of time scales is quite frequent in nature, while the tuning of parameters to a certain finite value only takes place accidentally. Thus, the forest–fire model is critical over a wide range of parameter values. A snapshot of the critical state is shown in Fig. 1.

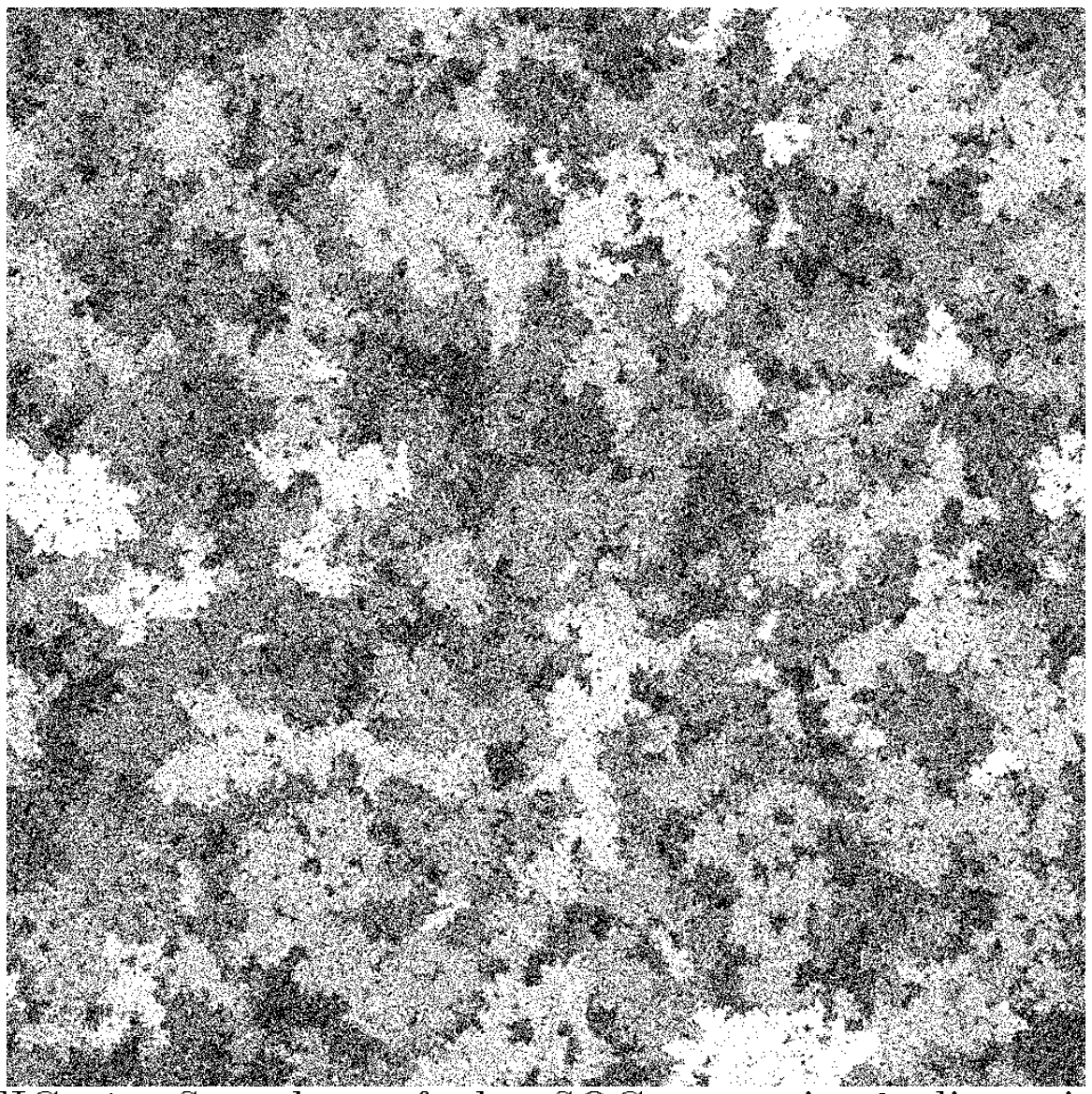

FIG. 1. Snapshot of the SOC state in 2 dimensions. Trees are black, empty sites are white. The parameters are $L = 1024$ and $f/p = 1/500$.

## III. SCALING LAWS AND CRITICAL EXPONENTS

In this section, we will derive scaling laws and relations between the critical exponents for the SOC forest–fire model.

First, we calculate the mean number $\bar{s}$ of trees that are destroyed by a lightning stroke. During one time step, there are

$$f\rho_t L^d$$

lightning strokes in the system and

$$p\rho_e L^d$$

growing trees. In the steady state, the number of growing trees equals the number of burning trees, and therefore the mean number of trees destroyed by a lightning stroke is

$$\bar{s} = \frac{p\rho_e}{f\rho_t} \simeq \frac{p}{f}\frac{1-\rho_t}{\rho_t}. \tag{3.1}$$

In the last step, we have neglected the fire density which is very small due to time scale separation. For small values of $f/p$, the forest density $\rho_t$ assumes a constant value. If this constant value is less than 1, the second factor on the right–hand side of Eq. (3.1) is also constant for small $f/p$, and Eq. (3.1) then represents a power law

$$\bar{s} \propto (f/p)^{-1}. \tag{3.2}$$

In $d \geq 2$ dimensions, the critical forest density

$$\rho_t^c = \lim_{f/p \to 0} \rho_t, \tag{3.3}$$

in fact, must be less than 1, as the following consideration indicates: If the critical forest density were $\rho_t^c = 1$ in $d \geq 2$ dimensions, $\rho_t$ would be very close to 1 for small values of $f/p$. Then the largest forest cluster would contain a nonvanishing percentage of all trees in the system, and the average number $\bar{s}$ of trees burned by a lightning stroke would diverge in the limit $L \to \infty$ with fixed $f/p$, in contradiction to Eq. (3.1). In one dimension, there is no infinitely large forest cluster in the system as long as $\rho_t < 1$, and therefore the critical forest density is $\rho_t^c = 1$. Nevertheless Eq. (3.2) holds also in 1 dimension since the forest density approaches its critical value only logarithmically slowly, as will be shown below. Eq. (3.2) indicates a critical point in the limit $f/p \to 0$. Close to this critical point, i.e. if $f \ll p$, there is scaling over many orders of magnitude.

Let $n(s)$ be the mean number of forest clusters per unit volume consisting of $s$ trees. Then the mean forest density is

$$\rho_t = \sum_1^\infty s n(s), \tag{3.4}$$

and the mean number of trees destroyed by a lightning stroke is

$$\bar{s} = \sum_1^\infty s^2 n(s)/\rho_t. \tag{3.5}$$

Since $\lim_{f/p \to 0} \rho_t$ is finite and $\bar{s}$ diverges $\propto (f/p)^{-1}$, these equations imply that $n(s)$ decreases at least like $s^{-2}$ but not faster than $s^{-3}$. As long as the system is not exactly at the critical point $f/p = 0$, i.e. for nonvanishing $f/p$, there must be a cutoff in the cluster size distribution for very large forest clusters. We conclude that [8]

$$n(s) \propto s^{-\tau} \mathcal{C}(s/s_{\max}) \tag{3.6}$$

with $2 \leq \tau \leq 3$ and

$$s_{\max}(f/p) \propto (f/p)^{-\lambda} \propto \bar{s}^\lambda. \tag{3.7}$$

The cutoff function $\mathcal{C}(x)$ is essentially constant for $x \leq 1$ and decreases to zero for large $x$. Eqs. (3.5) - (3.7) yield $\bar{s} \propto s_{\max}^{3-\tau}$, which leads to the scaling relation

$$\lambda = 1/(3-\tau). \tag{3.8}$$

In the case $\tau = 2$, the right–hand side of Eq. (3.6) acquires a factor $1/\ln(s_{\max})$ and reads now

$$n(s) \propto s^{-\tau} \mathcal{C}(s/s_{\max})/\ln(s_{\max}), \tag{3.9}$$

since the forest density given by Eq. (3.4) must not diverge in the limit $f/p \to 0$. The mean number of forest clusters per unit volume $\sum_1^\infty n(s)$, therefore, decreases to zero for $f/p \to 0$, and consequently the forest density approaches the value 1.

We also introduce the cluster radius $R(s)$ (radius of gyration) which is the mean distance of the trees in a cluster from their center of mass. It is related to the cluster size $s$ by

$$s \propto R(s)^\mu \tag{3.10}$$

with the fractal dimension $\mu$.

The pair connectedness $C(\mathbf{x}, f/p)$ is the probability that a site at distance $\mathbf{x}$ from an occupied site is also occupied and belongs to the same cluster [19]. The correlation length $\xi$ is defined by

$$\begin{aligned}
\xi^2 &= \frac{\sum_{\mathbf{x}} \mathbf{x}^2 C(\mathbf{x}, f/p)}{\sum_{\mathbf{x}} C(\mathbf{x}, f/p)} \\
&= \frac{\sum_1^\infty 2sn(s)\langle\sum_{i=1}^s R_i^2\rangle}{\sum_1^\infty sn(s)\sum_{i=1}^s 1} \\
&= \frac{\sum_1^\infty 2sn(s) \cdot sR^2(s)}{\sum_1^\infty sn(s) \cdot s} \\
&\propto (f/p) \int_1^\infty ds\, s^{2-\tau+2/\mu}\, \mathcal{C}(s/s_{max}) \\
&\propto (f/p)^{-2\lambda/\mu},
\end{aligned} \tag{3.11}$$

where $\langle \ldots \rangle$ denotes the average over all clusters of size $s$. We conclude

$$\xi \propto (f/p)^{-\nu} \text{ with } \nu = \lambda/\mu. \tag{3.12}$$

Another quantity of interest is the mean cluster radius

$$\bar{R} = \sum_{s=1}^\infty sn(s)R(s)/\sum_{s=1}^\infty sn(s)$$
$$\begin{cases} \propto (f/p)^{-(\nu-\lambda+1)}, & \text{if } \nu - \lambda + 1 \geq 0\,; \\ = \text{const.}, & \text{if } \nu - \lambda + 1 < 0\,. \end{cases}$$

This leads to

$$\bar{R} \propto (f/p)^{-\tilde{\nu}} \text{ with } \tilde{\nu} = \max(0, \nu - (\lambda - 1)). \tag{3.13}$$

In percolation theory, the *hyperscaling relation*

$$d = \mu(\tau - 1) \tag{3.14}$$

is satisfied, but it is not satisfied in the SOC forest–fire model in $d = 2$, as first stated in [12], where also an interpretation of this relation is given: If Eq. (3.14) is satisfied, every box of $l^d \gg 1$ sites contains a spanning piece of a large cluster when the system is at the critical point. In the forest–fire model, there are at least in $d = 2$ many regions which contain no large forest cluster (see Fig. 1), and consequently $d < \mu(\tau - 1)$.

The mean forest density $\rho_t$ approaches its critical value $\rho_t^c = \lim_{f/p\to 0} \rho_t$ via a power law

$$\rho_t^c - \rho_t \propto (f/p)^{1/\delta}. \tag{3.15}$$

One might try to calculate $1/\delta$ in the following way

$$\begin{aligned} \rho_t^c - \rho_t &\propto \sum_{s=1}^{\infty} s^{1-\tau} \left(1 - \mathcal{C}(s/s_{\max})\right) \\ &= \int_{s_{\max}}^{\infty} ds\, s^{1-\tau} \left(1 - \mathcal{C}(s/s_{\max})\right) \\ &\propto s_{\max}^{2-\tau} \propto (f/p)^{(\tau-2)/(3-\tau)}. \end{aligned} \tag{3.16}$$

The simulation results for $\tau$ and $\delta$, however, show that

$$(\tau - 2)/(3 - \tau)$$

is much smaller than $1/\delta$ (see Tab. I). We conclude that

$$\int_1^{\infty} dx\, x^{1-\tau} \left(\mathcal{C}(x) - 1\right) = 0 \tag{3.17}$$

and that the above formula Eq. (3.6) for $n(s)$ acquires an additional factor

$$1 - (f/p)^{1/\delta} \Delta n(s)\,.$$

Eq. (3.17) says that the number of trees in the system is not influenced by the cutoff function. All trees that were in large forest clusters if there were no cutoff can be found in smaller clusters. The size distribution of forest clusters therefore has a bump (see Fig. 2). This explanation of the bump has already been suggested in [10]. The function $\Delta n(s)$ determines how the distribution of forest clusters of size $s < s_{\max}$ deviates from the critical distribution. Since the mean density of empty sites decreases with decreasing $f/p$, less small forest clusters are produced when the system is closer to the critical point $f/p \to 0$. Consequently $\Delta n(s)$ is negative for small $s$. Since the forest density increases with decreasing $f/p$, the function $\Delta n(s)$ must be positive for larger values of $s$, leading to

$$\int_1^{\infty} ds\, s^{-\tau} \mathcal{C}(s/s_{max})\, \Delta n(s) > 0\,.$$

Finally, we introduce dynamical exponents characterizing the temporal behavior of the fire. Let $T(s)$ be the average time a cluster of size $s$ needs to burn down when ignited, and $N(T)$ the portion of fires that live exactly for $T$ time steps. Then the exponents $b$ and $\mu'$ are defined by

$$s \propto T(s)^{\mu'} \text{ and } N(T) \propto T^{-b}. \tag{3.18}$$

From

$$N(T)dT \propto sn(s)ds$$

follows the scaling relation

$$b = \mu'(\tau - 2) + 1. \tag{3.19}$$

The time scale of the system is set by

$$T_{\max} = T(s_{\max}) \propto (f/p)^{-\nu'} \text{ with } \nu' = \lambda/\mu'. \tag{3.20}$$

The dynamical critical exponent $z$ is defined by

$$T_{\max} \propto \xi^z,$$

which leads with (3.12) and (3.20) to

$$z = \nu'/\nu = \mu/\mu'. \tag{3.21}$$

The condition of time scale separation now can be expressed in terms of the critical exponents and reads

$$(f/p)^{-\nu'} \ll p^{-1} \ll f^{-1}, \tag{3.22}$$

or equivalently

$$f \ll p \ll f^{\nu'/(1+\nu')}. \tag{3.23}$$

The average lifetime of fires is

$$\bar{T} = \sum_{s=1}^{\infty} sn(s)T(s) / \sum_{s=1}^{\infty} sn(s) \propto (f/p)^{-\tilde{\nu}'} \tag{3.24}$$

with

$$\tilde{\nu}' = \max(0, \nu' - (\lambda - 1))\,. \tag{3.25}$$

The average number $N_s(t)$ of trees that burn $t$ time steps after a cluster of size $s$ is struck by lightning enters the definition of the temporal fire–fire correlation function $G(\tau)$

$$G(\tau) \propto \sum_{s=1}^{\infty} n(s) s \sum_{t=0}^{\infty} N_s(t) N_s(t+\tau). \tag{3.26}$$

The power spectrum is the Fourier transform of the fire–fire correlation function

$$G(\omega) = 2\int_0^{\infty} d\tau\, G(\tau)\cos(\omega\tau) \propto \omega^{-\alpha} \text{ for small } \omega. \tag{3.27}$$

Using the function $N_s(t)$, the mean chemical distance of the trees in a cluster from the site of the lightning stroke can be defined [20]:

$$\bar{t}(s) = \sum_{t=0}^{\infty} t N_s(t)\,. \tag{3.28}$$

This definition is analogous to the definition of the cluster radius as the mean distance of the trees of a cluster from the center of mass. When the structure of the forest clusters is compact or fractal, the chemical radius is proportional to the life time of the fire:

$$\bar{t}(s) \propto T(s)\,, \tag{3.29}$$

and $\bar{t}(s)$ introduces no new critical exponents. Consequently $T_{\max}$ is proportional to a chemical correlation length, which could be defined by a relation similar to Eq. (3.11). Our computer simulations confirm Eq. (3.29).

## IV. VALUES OF THE CRITICAL EXPONENTS IN 1 TO 8 DIMENSIONS

In this section, we determine the values of the critical exponents in 1 to 8 dimensions. In one dimension, the critical exponents can be determined analytically, as was done in [9]. In higher dimensions one has to resort to computer simulations. Here, we shortcut the exact evaluation of the critical exponents in $d = 1$ by using simple arguments. In subsection IV B we present the simulation results in 2 and more dimensions [21]. In subsection IV C we discuss the behavior of the system on a Bethe lattice.

### A. The critical exponents in one dimension

In one dimension, the critical forest density $\rho_t^c$ equals 1, since otherwise there were no infinitely large forest cluster in the system. The consideration after Eq. (3.8) shows that consequently $\tau = 2$ and (via scaling relation Eq. (3.8)) $\lambda = 1$. In the steady state, the density of forest clusters $\sum n(s)$ is constant, and therefore [22]

$$\sum_{s=1}^{\infty} n(s) = (1 - \rho_t - (f/p)\rho_t)/2\,,$$

which leads together with Eq. (3.9) to

$$(1-\rho_t) \propto 1/\ln(s_{\max})$$

and $1/\delta = 0$. One–dimensional forest clusters are compact, therefore $\mu = 1$ and (with Eq. (3.12)) $\nu = 1$. From $T(s) \propto R(s)$ it follows $\mu' = \nu' = z = 1$. The exact calculation in [9] yields additional logarithmic corrections:

$$s_{\max} \propto \xi \propto T_{\max} \propto (p/f)/\ln(p/f).$$

The fourier transform of the temporal correlation function is [9]

$$G(\omega) \propto \omega^{-2}(1 + \mathrm{const.}\cdot \ln(\omega s_{\max}))\,, \tag{4.1}$$

indicating a deviation from the trivial $\omega^{-2}$-dependence towards $1/\omega$–noise. Tab. I summarizes the values of the critical exponents. They are confirmed by our simulations.

### B. Simulation results in 2 and more dimensions

We obtained the values of the critical exponents in $d \geq 2$ dimensions by computer simulations using the same method as in [10] which iterates the following rules:

1. Choose an arbitrary site in the system. If it is not occupied by a tree, proceed with rule 2. If it is occupied by a tree, then ignite the tree and burn down the forest cluster to which the tree belongs. While burning the trees, evaluate the size and the radius of the cluster, the lifetime of the fire and the function $N_s(t)$. Proceed with rule 2.

2. Choose $p/f$ arbitrary sites in the system and grow a tree at all chosen empty sites. Proceed with rule 1.

Time scale separation is perfectly realized by these rules, and Eq. (3.1) is satisfied.

In order to assure that the system is in the steady state, a sufficiently large number of time steps have been discarded in the beginning of each simulation. The simulations were performed on a DECstation 5000. The system length $L$ for each dimension used in our simulations is given in Tab. I. The simulations could not be performed arbitrarily close to the critical point $f/p \to 0$, since the correlation length $\xi$ diverges, and, therefore, finite–size effects occur when $\xi$ exceeds the system size. Especially the distance–related exponents $\mu$, $\nu$ and $\tilde{\nu}$ could not be determined with available computer capacity in high dimensions. In the following, we discuss the simulation results obtained for the critical exponents and the critical forest density.

*a. Exponent $\tau$* In contrast to the rest of the critical exponents, the exponent $\tau$ can also be determined in higher dimensions, since the cluster distribution "feels" more the overall system size $L^d$ than the edge length $L$. Therefore the exponent $\tau$ could be measured in 1 to 8, 12 ($L = 4$) and 16 ($L = 3$) dimensions (see Fig. 2). Since we evaluated the forest clusters struck by lightning, we measured the fire distribution $sn(s)$, which gives $\tau - 1$.

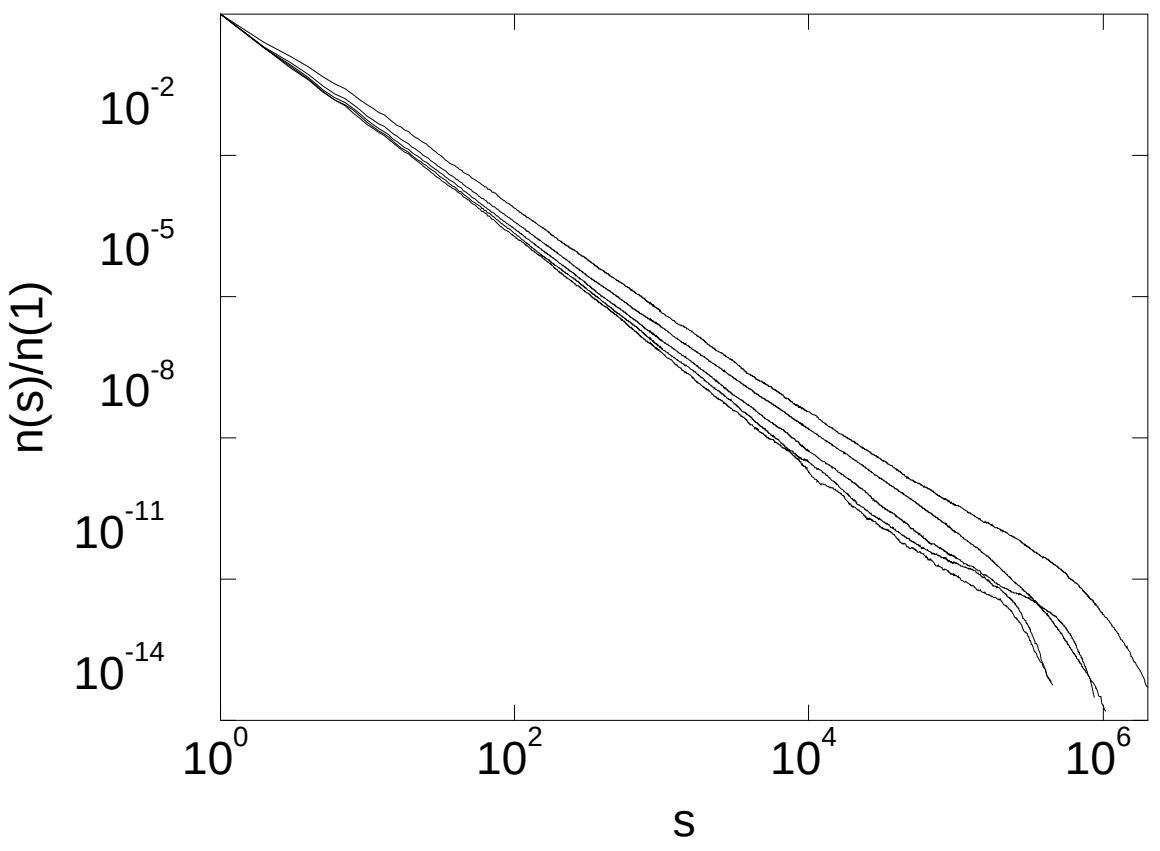

FIG. 2. Normalized cluster distribution $n(s)/n(1)$ in $d=2$ to 6 dimensions. The values of $f/p$ are 1/32000, 1/2000, 1/1000, 1/250 and 1/125 from right to left. The negative slope yields the exponent $\tau$.

The values of $f/p$ which just don't lead to finite–size effects are determined by means of the *integrated fire distribution* $P(s)$, which is defined [10] by

$$P(s) = \int_s^\infty ds'\, s' n(s'). \tag{4.2}$$

A plot of $P(s)/P(1)$ for different values of $f/p$ in $d=6$ is shown in Fig. 3.

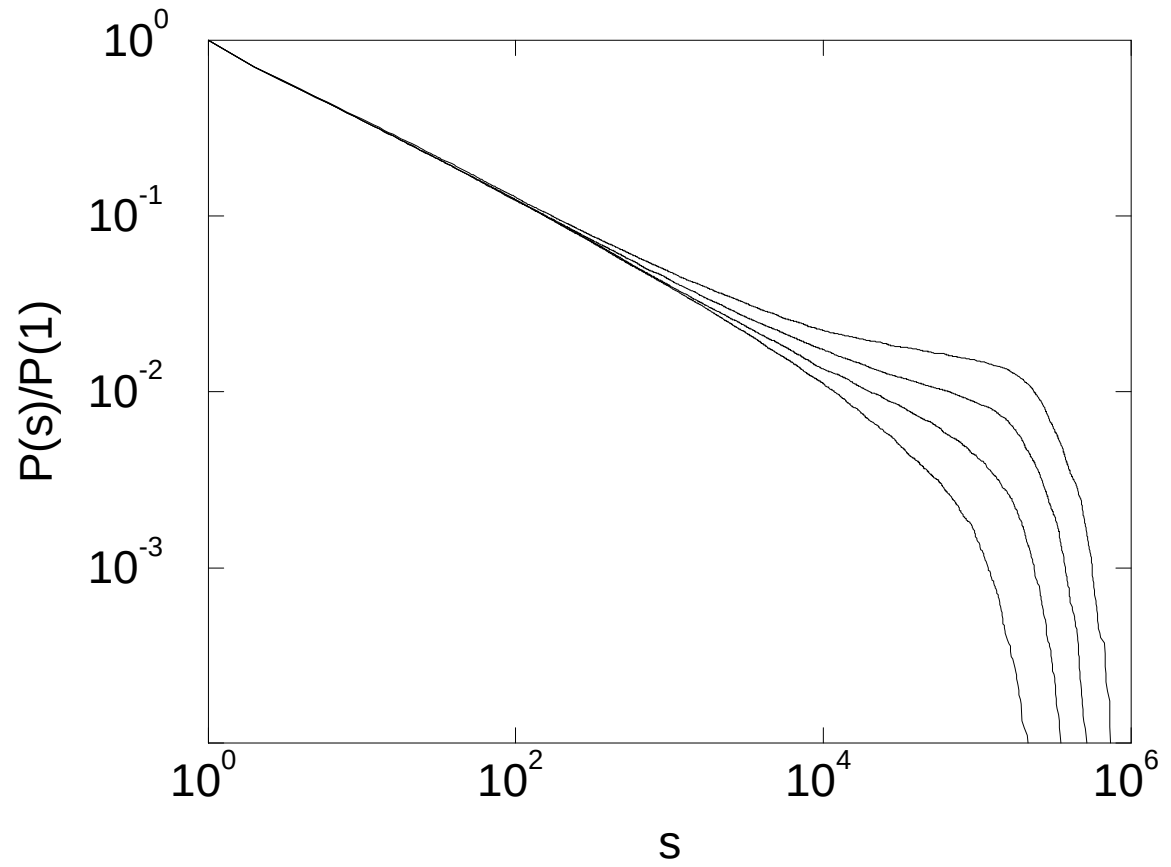

FIG. 3. Normalized integrated fire distribution $P(s)/P(1)$ for $f/p = 2/125$, 1/125, 1/250, 1/500 in $d=6$ (from left to right).

As long as the finite size of the system is not important, the scaling regime of $P(s)$ becomes larger, if one approaches the critical point. As soon as $P(s)$ feels the finite system size, the curves bend upwards (see the two uppermost curves in Fig. 3). Evaluating the curves not affected by finite–size effects, we obtain the values of $\tau$ shown in Tab. I. For $d=2$, they are consistent with [10,12]. The values of $\tau$ for $d=2$ to 6 are compatible with those given in [11] but have a smaller error since they have been obtained using larger systems closer to the critical point. Above $d=6$ dimensions, the numerical results suggest $\tau = 2.5$ as in percolation theory. Our results indicate that the SOC forest–fire model has the *upper critical dimension* $d_c = 6$, as already conjectured in [10,11]. An analytic proof, however, is still missing.

*b. Exponent* $\lambda$ The determination of $\lambda$ is slightly more difficult. Consider again the normalized integrated distribution function $P(s)/P(1)$:

$$\begin{aligned}
\ln \frac{P(s)}{P(1)} &= \ln \frac{\int_s^\infty ds'\, s'^{1-\tau} \mathcal{C}(s'/s_{\max})}{\int_1^\infty ds'\, s'^{1-\tau} \mathcal{C}(s'/s_{\max})} \\
&= \ln \int_{s/s_{\max}}^\infty dx\, x^{1-\tau} \mathcal{C}(x) - \ln \int_{1/s_{\max}}^\infty dx\, x^{1-\tau} \mathcal{C}(x) \\
&= A(s/s_{\max}) - A(1/s_{\max}).
\end{aligned} \tag{4.3}$$

Eq. (4.3) implies that the plots of $\ln(P(s)/P(1))$ vs. $\ln(s)$ for different values of $f/p$ can be made to coincide, when they are shifted horizontally and vertically (see Fig. 4).

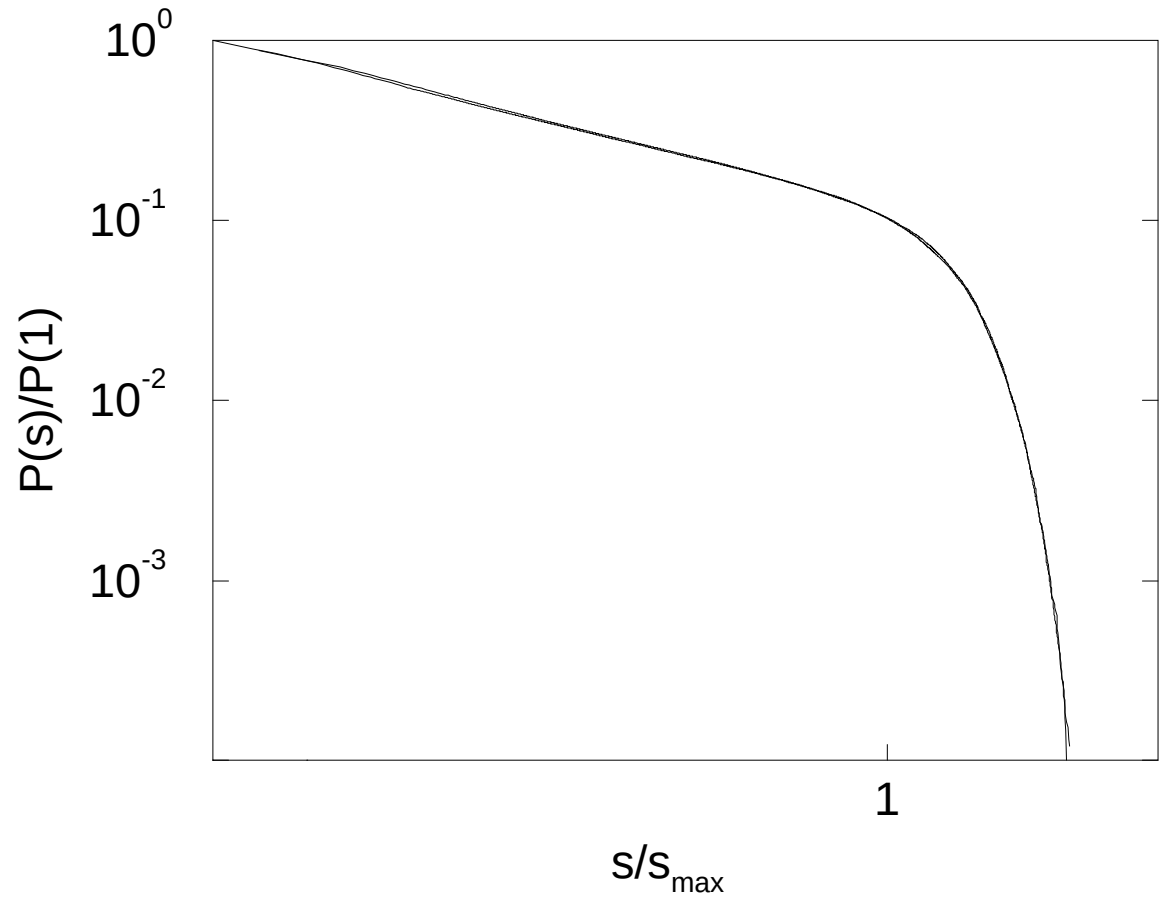

FIG. 4. Normalized integrated fire distribution $P(s)/P(1)$ for $f/p = 1/16000$ and $f/p = 1/32000$ in $d=2$.

The logarithm of the horizontal scaling factor divided by $\ln((f/p)_1/(f/p)_2)$ then yields $\lambda$. Our simualtion results in $d=2$ and 3 are given in Tab. I. They satisfy the scaling relation Eq. (3.8).

With increasing dimension, the range of admissible values of $f/p$ decreases due to finite–size effects. Consequently, the error in $\lambda$ becomes fairly large in $d \geq 4$. Hence, in these dimensions, its value was calculated using the scaling relation Eq. (3.8).

For $d=2$, the value of $\lambda$ has also been determined in [12] which agrees with ours, and in [10] which does not agree with ours and is inconsistent with the scaling relation Eq. (3.8).

*c. The critical forest density* $\rho_t^c$ *and the exponent* $\delta$ In our simulations the forest density $\rho_t$ was determined by dividing the number of burnt–down clusters by the number of iterations. In order to calculate the critical forest density $\rho_t^c$ and the exponent $\delta$, one needs at least three simulations at different $f/p$. In $d=2,3$ we fitted the results $\rho_t(f/p)$ with the function $\rho_t^c - \text{const.} \cdot (f/p)^{1/\delta}$ (Fig. 5).

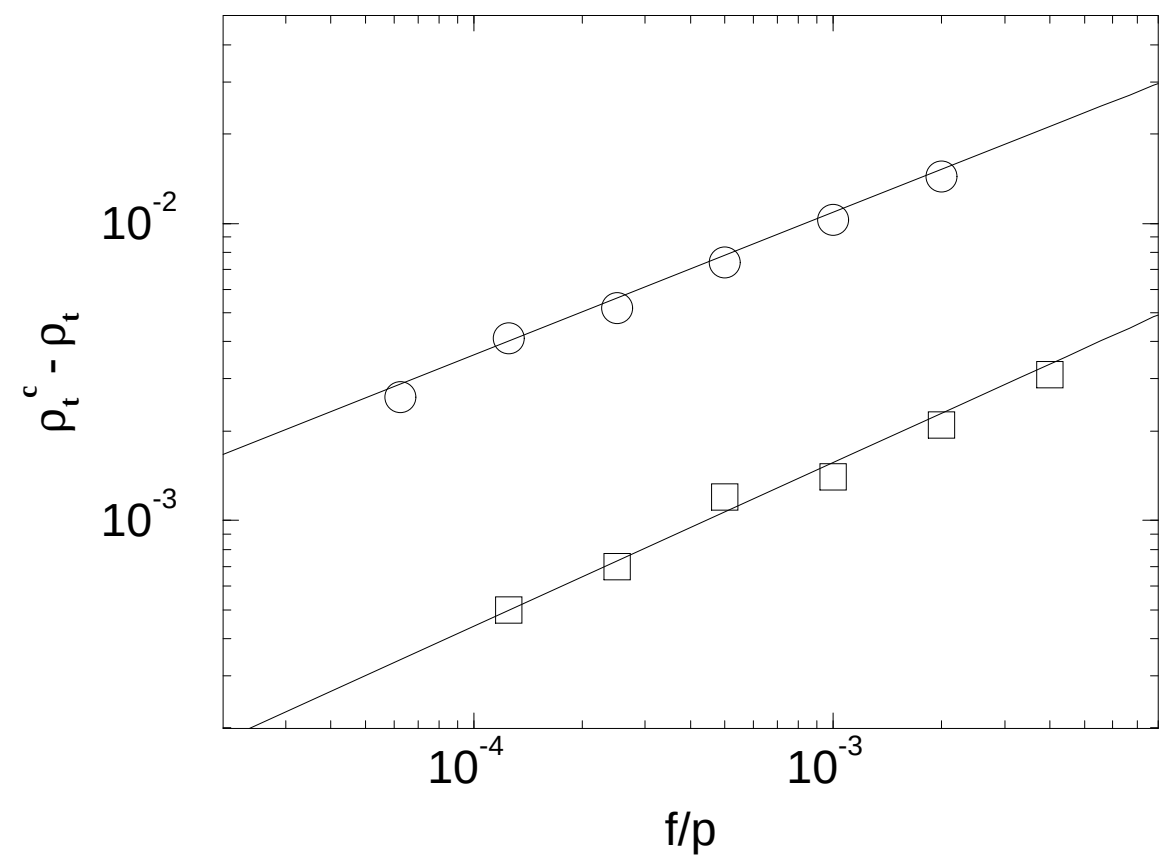

FIG. 5. $\rho_t^c - \rho_t$ in $d = 2$ (∘) and $d = 3$ (□). The inverse slope yields the exponent $\delta$.

The error in $1/\delta$ is small enough to rule out the possibility that $1/\delta = (\tau - 2)/(3 - \tau)$ (see Eq. (3.16)). In $d > 3$ the change in $\rho_t(f/p)$ and the error in $\rho_t$ are of the same order of magnitude, making the fit useless.

The results for $\rho_t^c$ and $1/\delta$ as well as $1/\delta$ in percolation theory and $(\tau - 2)/(3 - \tau)$ are given in Tab. I. Also shown in Tab. I are the corresponding values of the mean–field forest density. In mean–field theory the critical forest density equals $1/(2d - 1)$, when taking into account that burning trees always have an empty neighbor where the fire has come from. $\rho_t^c$ is indistinguishable from the mean–field density for $d \geq 5$.

Our value of $\rho_t^c$ in $d = 2$ is consistent with [10,11]. The result in [12] for $d = 2$ as well as the results in [11] for $d = 3$ to 6 are moderately larger than ours, which is probably due to the smaller system size of these other simulations.

Our values of $1/\delta$ are compatible with [10,11], whereas the result in [12] for $d = 2$ seems to be too small. In three dimensions our result for $\delta$ is already very close to the percolation value.

*d. Exponents $\mu$, $\nu$ and $\tilde{\nu}$* The fractal dimension $\mu$ of the forest clusters is obtained from the slope of the cluster radius $R(s)$ (see Fig. 6).

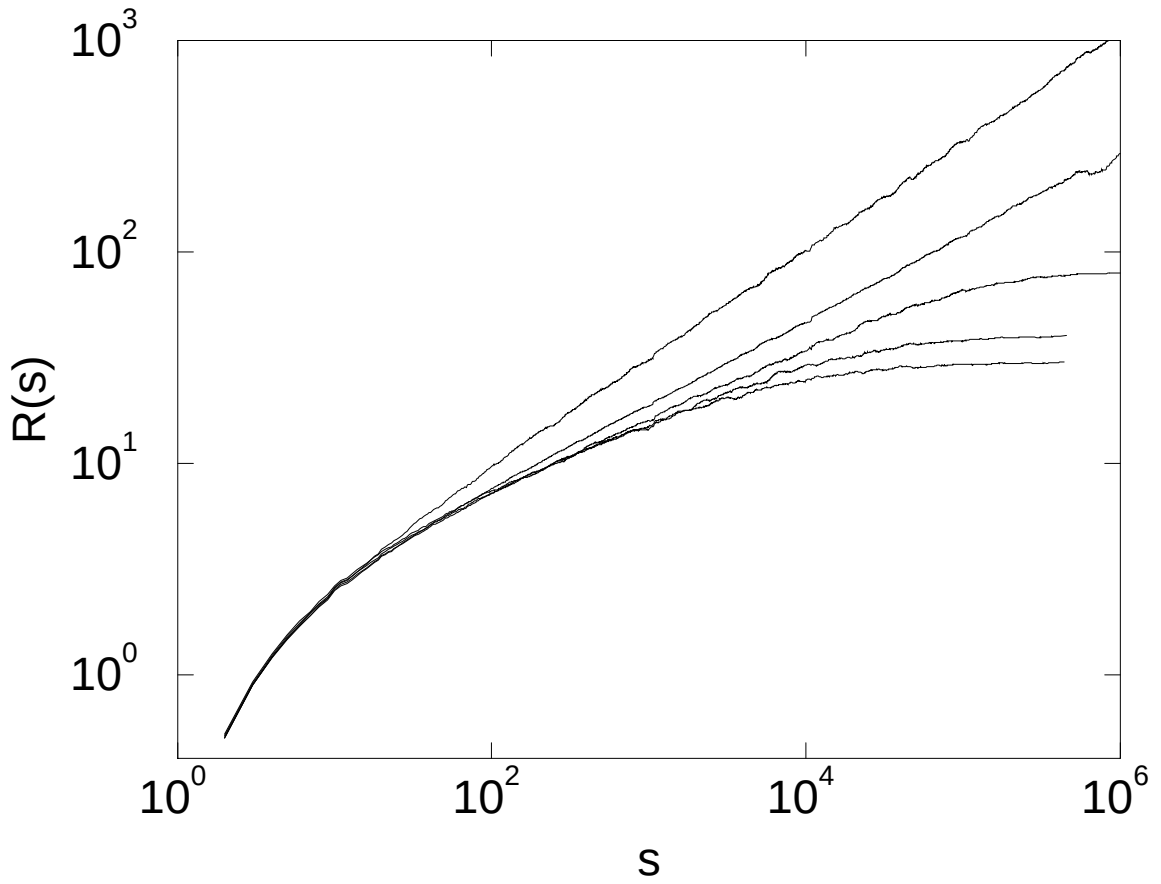

FIG. 6. Cluster radius $R(s)$ in 2 to 6 dimensions (from left to right). The inverse slope yields the fractal dimension $\mu$.

The distance between two lattice sites $\mathbf{x}, \mathbf{x}'$ is defined by

$$d(\mathbf{x}, \mathbf{x}') = \sum_{i=1}^{d} |x_i - x_i'| \,, \tag{4.4}$$

since this is the length of the shortest path between them. In 6 and higher dimensions the $R(s)$–plot has no scaling region, therefore $\mu$ could only be determined in $d = 2$ to 4 and, with relatively large error, in $d = 5$. Unlike [8,10], we find $\mu < d$ in $d = 2$, in accordance with [12]. From the results in Tab. I it seems that $\mu$ approaches the fractal dimension of percolation clusters and we suggest that $\mu = \mu_{\text{perc}}$ for $d \geq 6$. The hyperscaling relation Eq. (3.14) is definitely violated in $d = 2$, but holds within error margins in $d = 3, 4, 5$.

The correlation length is dominated by large clusters and consequently large radii. Therefore the exponent $\nu$ could directly be determined only in $d = 2$ dimensions. Since $\bar{R}$ is less divergent than $\xi$, the exponent $\tilde{\nu}$ could also be determined in $d = 3$ dimensions. Our value of $\nu$ in $d = 2$ is compatible with [10,12], and all results (see Tab. I) satisfy the scaling relations Eqs. (3.12) and (3.13).

*e. Exponents $\mu'$, $\nu'$, and $\tilde{\nu}'$* The *chemical dimension* $\mu'$ [20] can be determined even in $d = 16$, since the scaling regime extends over at least two decades. In 3 and more dimensions, the results are compatible with the mean–field value $\mu' = 2$. Just as $\nu$ and $\tilde{\nu}$, the exponents $\nu'$ and $\tilde{\nu}'$ could only be determined in $d = 2$ and $d = 2, 3$ respectively. The results (see Tab. I) obey the scaling relations Eqs. (3.25) and (3.20).

*f. Exponents $b$ and $z$* Knowing the exponents $\mu$, $\mu'$ and $\tau$, one can easily calculate the exponent of the distribution of fire lifetimes $b$ (see Eq. (3.19)) and the dynamical exponent $z$, which characterizes the relation between the time scale and the length scale of the system (see Eq. (3.21)). The results are shown in Tab. I. Both exponents seem to approach the value 2 for $d \to 6$.

*g. Exponent $\alpha$* Assuming a simplified law $N_s(t) \propto t^{\mu'-1}\,\Theta(T(s) - t)$, one can show that (to leading order in the frequency) $\alpha = 2$ for $d \geq 6$. For $1 < d < 6$, $\alpha$ has to be determined by simulations. The results are shown in Tab. I. The non–monotonous behaviour of $\alpha$ between $d = 1$ and $d = 6$ is not so surprising if one recalls the intricate dependence of $\alpha$ on $\mu'$ and $\tau$ (see Eq. (3.26)).

In summary, we can conclude that the SOC forest-fire model is likely to have an upper critical dimension $d_c = 6$, above which the critical exponents are identical with those of mean–field–theory, which again is identical to the mean–field–theory of percolation. The strongest evidence for this behaviour comes from the exponent $\tau$, which approaches the percolation value $\tau_{\text{perc}} = 5/2$ for $d \to 6$ and is indistinguishable from 5/2 in all simulated dimensions $d \geq 6$. But also the difference in the other exponents between forest–fire and percolation values seems to vanish with increasing dimension. Of course a non integer upper critical dimensionality as $d_c = 11/2$ would also be compatible with our results.

### C. The SOC forest–fire model on the Bethe lattice

Often the critical exponents of a model in high dimensions are obtained not only by mean–field theory but also on the Bethe lattice. This might also be true for the SOC forest–fire model. The tree distribution on the Bethe lattice, however is not random and therefore different from mean–field theory, as the following consideration indicates:

A forest cluster of $s$ trees on a Bethe lattice with configuration number $z$ has

$$s(z-2)+2$$

empty neighbors, independently of its form. It is destroyed either when a tree grows at its edge (this happens with probability $p\,(s\,(z-2)+2)$ per time step), or when it is struck by lightning (with probability $fs$). A new cluster is generated each time a tree grows. In the steady state, the mean number of forest clusters is constant, i.e. [22]

$$\begin{aligned} p\,(1-\rho_t) &= \sum_{s=1}^{\infty}(fs + p\,(2+s\,(z-2)))n(s) \\ &= \rho_t\,(f+p\,(z-2)) + 2p\sum_{s=1}^{\infty} n(s)\,. \end{aligned} \tag{4.5}$$

In the limit $f/p \to 0$ it follows from Eq. (4.5)

$$\sum_{s=1}^{\infty} n(s) = (1-\rho_t\,(z-1))/2\,. \tag{4.6}$$

If the stationary critical state were identical with mean–field theory of the forest–fire model, which is known to be identical with the mean–field theory of percolation [11,23], i.e. with percolation on a Bethe lattice, trees were randomly distributed with

$$\rho_t = 1/(z-1)\,.$$

With Eq. (4.6) this would lead to

$$\sum_{s=1}^{\infty} n(s) = 0\,,$$

i.e. the density of forest clusters would decrease to zero in the limit $f/p \to 0$. Consequently, the exponent $\tau$ were 2 (see the comment after Eq. (3.8)). Both results are in contradiction to the well–known facts that a random tree distribution with density $1/(z-1)$ leads to a finite density of clusters and to $\tau = 2.5$ on the Bethe lattice. Therefore the SOC forest–fire model on the Bethe lattice has no random distribution of trees and is different from its mean–field–theory. But we cannot rule out the possibility that the critical exponents nevertheless are the same as in mean–field theory. For the sandpile model on the Bethe lattice, it has been shown that the asymptotic power laws are the same as in mean–field theory, although there exist short-range correlations between sites [24].

## V. UNIVERSALITY OF THE CRITICAL EXPONENTS

The critical behavior of a system usually depends only on properties as dimension and conservation laws, but not on microscopic details. We therefore expect that the critical exponents of the SOC forest–fire model are universal under certain changes of the model rules.

In order to check this assumption, we repeated the 2D simulations for other lattice types. First, we chose a triangular lattice. This is equivalent to a square lattice with fire spreading to next–nearest–neighbors along one of the diagonals. Then, we investigated the model on a square lattice with next–nearest–neighbor interaction.

The simulations of both variations of the model were done on a $4096 \times 4096$–lattice with $f/p$ ranging from 1/1000 to 1/8000. We compared the exponents $\tau$, $\mu$, $\mu'$, $\nu$, $\nu'$ and $\alpha$ with the results given in Tab. I and found them to be exactly the same (see Figs. 7,8,9,10).

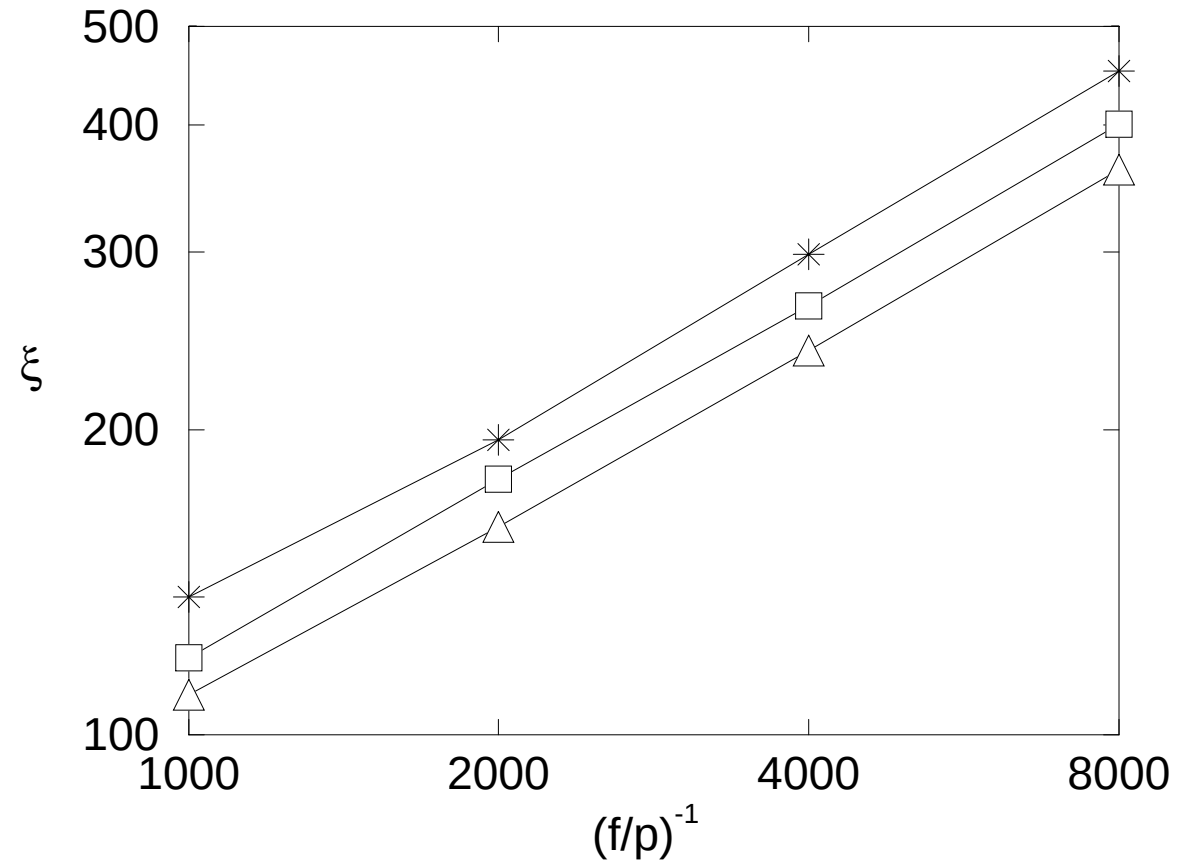

FIG. 7. The correlation length $\xi$ as function of $(f/p)^{-1}$. The slope yields the critical exponent $\nu$. (□ = square lattice, △ = triangular lattice, ∗ = next–nearest–neighbour interaction.)

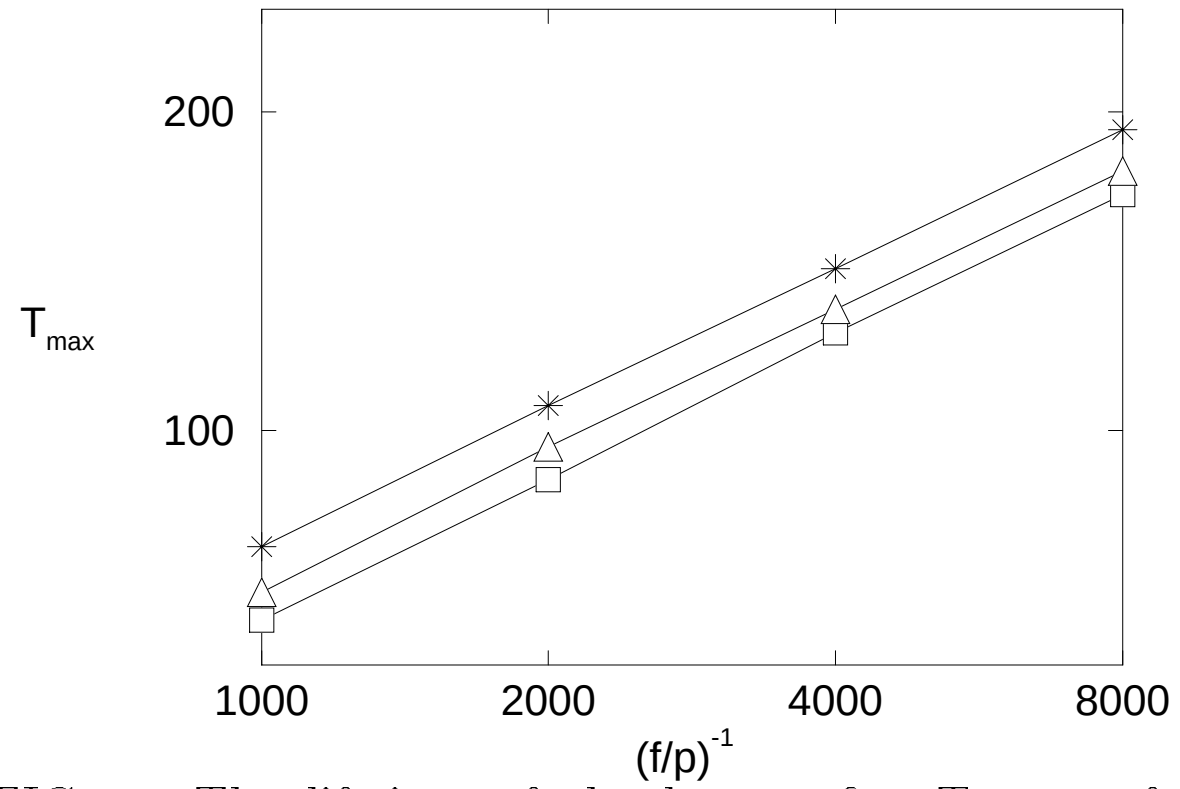

FIG. 8. The lifetime of the largest fire $T_{\max}$ as function of $(f/p)^{-1}$. The slope yields the critical exponent $\nu'$. (□ = square lattice, △ = triangular lattice, ∗ = next–nearest–neighbour interaction.)

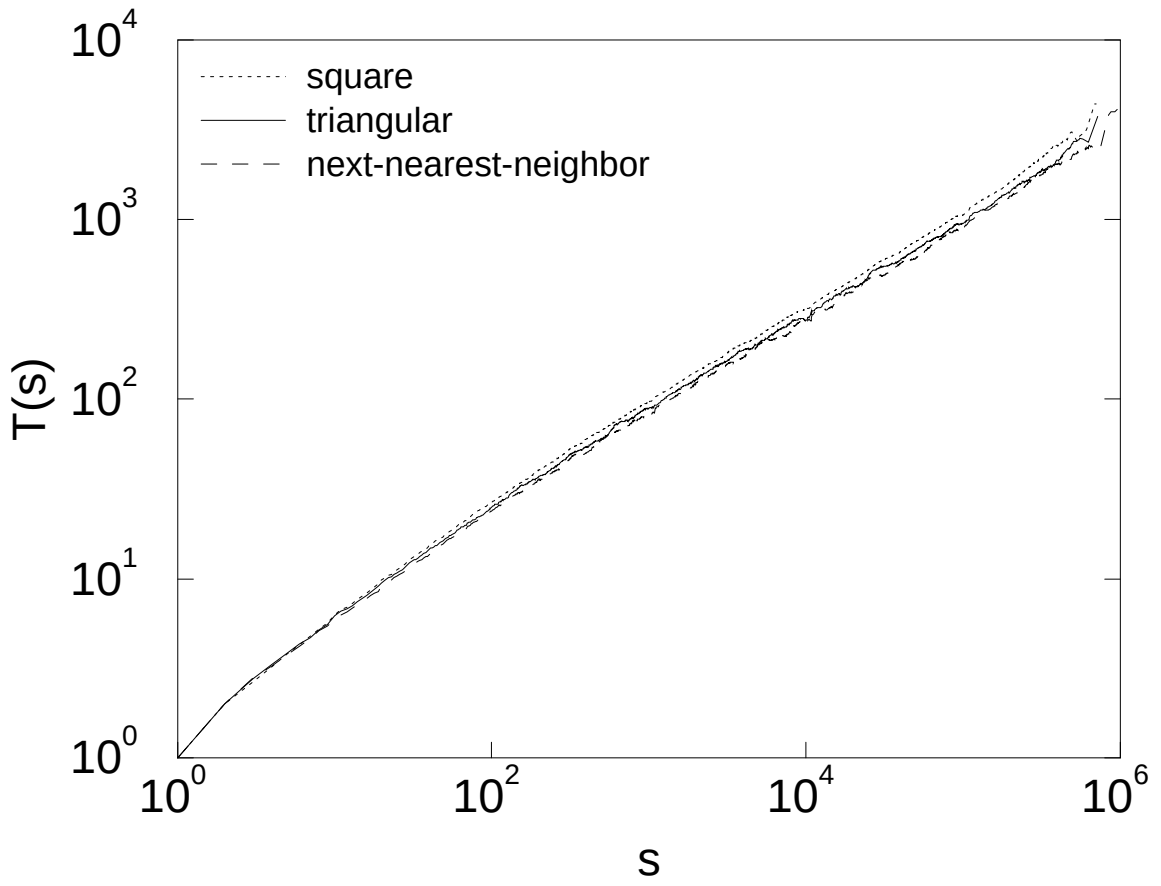

FIG. 9. Fire lifetime $T(s)$ for various lattice symmetries. The inverse slope yields the chemical dimension $\mu'$.

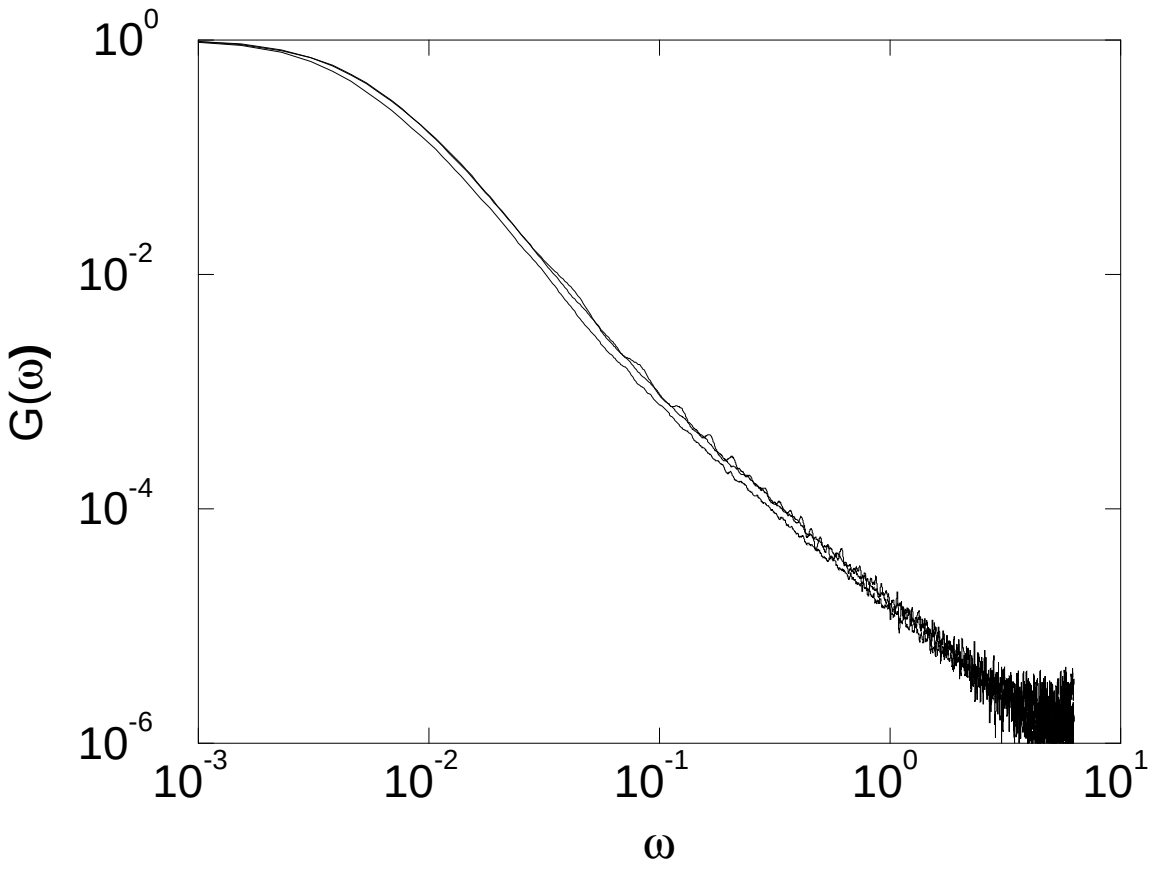

FIG. 10. The temporal correlation function $G(\omega)$ for various lattice symmetries. The negative slope yields the critical exponent $\alpha$.

Looking at Eq. (3.1) one expects the forest density for the triangular lattice to drop below the square–lattice value of $\rho_t^c \approx 40.8\%$, and the forest density for next–nearest–neighbor interaction to drop even further, since the fire has now more possible paths to spread. The values obtained from our simulations are $\rho_t^c \approx 34\%$ for the triangular lattice and $\rho_t^c \approx 28\%$ for next–nearest–neighbor interaction.

Another modification of the model rules is obtained when trees are allowed to be immune against fire. A forest–fire model with immune trees, but without lightning was investigated in [13]. Now we included the immunity in the SOC forest–fire model by changing rule 2 in the following way:

- tree $\longrightarrow$ burning tree with probability $1 - g^n$, if $n$ nearest neighbors are burning.

A detailed account on immunity in the SOC forest–fire model will be given elsewhere [25]. Here, we just state the main result: If one increases the immunity $g$ from 0 to some finite value below the critical value $g_c = 1/2$, which is one minus the bond–percolation threshold, the critical exponents remain unchanged. In order to see the asymptotic critical behaviour in the simulation data however, one has to go to very small values of $f/p$, when $g$ is close to $g_c$. For $g = 1/2$, the SOC fixed point disappears, and the size distribution of fires is identical to the size distribution of percolation clusters in uncorrelated bond percolation.

## VI. SUMMARY AND DISCUSSION

In this paper, we have presented a scaling theory for the SOC forest–fire model, including also the dynamics of the system. The appropriate critical exponents were defined, and scaling relations between them were derived.

The critical exponents in one dimension, which are known exactly [9], were rederived by simple arguments. They turned out to be just the classical ones proposed in [8]. In dimensions $\geq 2$, computer simulations then determined the values of the critical exponents and confirmed the scaling relations. Our simulations have been performed closer to the critical point than any earlier simulation. The values of many exponents have been determined for the first time. Results which have already been obtained by other authors are mostly confirmed, some of them are corrected or improved.

The simulations suggest that the critical exponents of the SOC forest–fire model in dimensions $d \geq d_c = 6$ are given by its mean–field theory, which is identical with the mean–field theory of percolation. It still remains a challenge to give an analytic derivation of the upper critical dimension. A short calculation showed that the SOC forest–fire model on the Bethe lattice is different from the mean–field theory of the model.

Finally, we investigated the universality of the critical properties by changing the lattice symmetry from "square" to "triangular" and by including next–nearest–neighbor interaction. Furthermore, the new parameter "immunity" was introduced into the SOC version of the forest–fire model. No violation of universality could be found.

As already pointed out in earlier publications [13,14], there is a close relationship between the forest–fire model and excitable media which comprise phenomena so different as spreading of deseases, oscillating chemical reactions, propagation of electrical activity in neurons or heart muscles, and many more (For a review on excitable systems see e.g. [26,27]). These systems essentially have three states which are called quiescent ($\triangleq$ tree), excited ($\triangleq$ burning tree), and refractory ($\triangleq$ empty site). Excitation spreads from one place to its neighbors if they are quiescent. After excitation, a refractory site needs some time to recover its quiescent state. In many of these systems, spiral–waves have been observed. We expect that a SOC state can be found in some of these systems, if the appropriate range of parameter values is investigated, i.e.

if spontaneous excitation occurs rarely and if excitation spreads much faster than the system recovers from the refractory state.

---

TABLE I. Numerical results for the critical exponents in 1 to 8 dimensions ($^*$ = with logarithmic corrections, $^\dagger$ = calculated from scaling relations). The exponents with index "perc" are those of percolation theory [19].

| $d$ | 1 | 2 | 3 | 4 | 5 | 6 | 7 | 8 |
|---|---|---|---|---|---|---|---|---|
| $L$ | $2^{20}$ | 16384 | 448 | 80 | 32 | 20 | 12 | 8 |
| $\tau$ | 2 | 2.14(3) | 2.23(3) | 2.36(3) | 2.45(3) | 2.50(3) | 2.50(3) | 2.50(3) |
| $\tau_{\rm perc}$ | 2 | 2.05 | 2.18 | 2.31 | 2.41 | 2.5 | 2.5 | 2.5 |
| $\lambda$ | $1^*$ | 1.15(3) | 1.30(6) | $1.56(8)^\dagger$ | $1.82(10)^\dagger$ | $2.01(12)^\dagger$ | $2.01(12)^\dagger$ | $2.01(12)^\dagger$ |
| $1/\delta$ | $0^*$ | 0.48(2) | 0.55(12) | - | - | - | - | - |
| $1/\delta_{\rm perc}$ | 1 | 0.42 | 0.56 | 0.69 | 0.85 | 1 | 1 | 1 |
| $\frac{\tau-2}{3-\tau}$ | 0 | 0.18 | 0.30 | 0.56 | 0.82 | 1 | 1 | 1 |
| $\rho_t^c$ | 1 | 0.4081(7) | 0.2190(6) | 0.146(1) | 0.111(1) | 0.090(1) | 0.076(1) | 0.066(1) |
| $\frac{1}{2d-1}$ | 1 | 0.333 | 0.200 | 0.143 | 0.111 | 0.091 | 0.077 | 0.067 |
| $\mu$ | 1 | 1.96(1) | 2.51(3) | 3.0 | 3.2(2) | - | - | - |
| $\mu_{\rm perc}$ | 1 | 1.90 | 2.53 | 3.06 | 3.54 | 4 | 4 | 4 |
| $\nu$ | $1^*$ | 0.58 | $0.52(3)^\dagger$ | $0.53(3)^\dagger$ | $0.57(7)^\dagger$ | - | - | - |
| $\tilde{\nu}$ | $1^*$ | 0.43 | 0.25 | $< 0.02^\dagger$ | $0^\dagger$ | $0^\dagger$ | $0^\dagger$ | $0^\dagger$ |
| $\mu'$ | 1 | 1.89(3) | 2.04(10) | 2.02(10) | 1.98(10) | 1.94(10) | 1.92(10) | 1.93(10) |
| $\mu'_{\rm perc}$ | 1 | 1.68 | 1.89 | | | 2 | 2 | 2 |
| $\nu'$ | $1^*$ | 0.58 | $0.64(6)^\dagger$ | $0.78(8)^\dagger$ | $0.92(10)^\dagger$ | $1.04(11)^\dagger$ | $1.05(11)^\dagger$ | $1.05(11)^\dagger$ |
| $\tilde{\nu}'$ | $1^*$ | 0.44 | 0.34 | $0.21(8)^\dagger$ | $0.1(1)^\dagger$ | $< 0.14^\dagger$ | $< 0.08^\dagger$ | $< 0.08^\dagger$ |
| $z$ | 1 | $1.04(2)^\dagger$ | $1.24(8)^\dagger$ | $1.49(10)^\dagger$ | $1.62(18)^\dagger$ | - | - | - |
| $b$ | 1 | $1.27(7)^\dagger$ | $1.47(9)^\dagger$ | $1.73(10)^\dagger$ | $1.89(11)^\dagger$ | $1.97(11)^\dagger$ | $1.96(11)^\dagger$ | $1.97(11)^\dagger$ |
| $\alpha$ | $2^*$ | 1.72(5) | 2.15(5) | 2.00(5) | 2.01(5) | 1.95(10) | 1.96(10) | 1.96(10) |